\documentclass[american,aps,prl,showpacs,twocolumn]{revtex4}
\usepackage{graphicx}

\makeatletter

\usepackage{babel}
\makeatother
\begin{document}

\title{The return branch of viscous fingers}

\author{Francisco Vera}

\email{fvera@ucv.cl}

\author{Mauricio Echiburu}

\affiliation{Pontificia Universidad Cat\'{o}lica de Valpara\'{\i}so, Av. Brasil
2950, Valpara\'{\i}so, Chile }

\begin{abstract}
We report a simple experiment of two-dimensional pattern formation
in a circular Hele-Shaw cell, showing the appearance of a return branch
that is equivalent to the upward-connecting leader of lightning. Injecting
water from the center into a foam filled cell, we obtained patterns
similar to dendrites of two-dimensional dielectric breakdown experiments.
When we repeat this experiment allowing the presence of water in the
outer (low pressure) region, dendrites
grow initially as in a normal experiment, but when a branch is near
the outer boundary, the low pressure water begins to penetrate the
foam against the pressure field, forming several return branches.
\end{abstract}


\maketitle

Typical experiments in a two-dimensional region between two parallel
plates or Hele-Shaw cell\cite{Hele-Shaw}, consist of injecting a
less viscous fluid into a more viscous fluid. These systems develop
patterns that were discovered for the first time when water was injected
to extract the remaining oil deposited in the bottom of natural wells.
Contrary to what was expected, water invaded the oil forming patterns
and frustrated the extraction. It is very easy to produce these patterns
in the laboratory, and many experiments have been reported for many
different geometries and fluids\cite{McCloud-Maher1995}. We report
a new class of patterns obtained in a radial geometry Hele-Shaw cell, injecting water
from the center into a foam that fills the two-dimensional region
between the plates. The novel ingredient is to allow the presence
of water at atmospheric pressure at the outer boundary.

In thunderstorms\cite{uman1989,uman2003}, air currents separate negative
and positive charges producing regions of high electric fields, when
this field exceeds the value for breaking the air, a lightning stroke
is produced. A typical negative cloud to ground stepped leader has
an overall duration of 35 ms and a current between 100-200 A. As the
tip of this negative leader get close to the earth surface, one or
more upward-moving discharges are initiated from the ground, and an
attachment with the downward-moving leader occurs some tens of meters
above ground. In this work, we obtained the equivalent of this effect
in a simple experiment using fluids under pressure, this experiment
shows that the \emph{return branch} is universal, and provides an
essential step towards the quest for an unified mechanism that explains
the basic features of different pattern forming systems. To our knowledge,
no experiment has ever been reported showing the appearance of a return
branch in a system different from lightning.

There are previous experimental works showing the appearance of universal
patterns in completely different systems: The Couder bubble\cite{couder1986,McCloud-Maher1995}
provided the mechanism for patterns typically formed in solidification\cite{mullins-sekerka,langer1980,biloni-boettinger},
to be found in experiments injecting a less viscous fluid towards
a more viscous one in a Hele-Shaw cell; Carving a lattice in
one of the plates of a Hele-Shaw cell, produce dendritic patterns
that look like snow crystals\cite{benjacob1985-grooves}; Bacterial
colonies in a Petri dish\cite{benjacob1993-snowtobacterial,benjacob2000-microorg}
under some conditions of nutrients and hardness of the gel, produce
beautiful patterns resembling fractals and spirals. Nature also provides
many examples of patterns showing some kind of universality\cite{ball-tapestry,witten-sander1981,cross-hohenberg1993}:
The blood vessels in our retina, the form of plants and simple animals,
the fracture patterns in solids, dielectric breakdown, the river networks,
etc.

\begin{figure}
\includegraphics[%
  width=1.0\columnwidth]{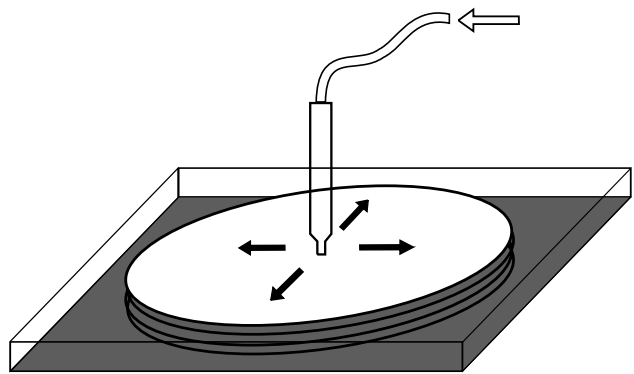}

\caption{Schematic view of the experimental setup}
\end{figure}

Motivated by the possibility for the existence of an unified mechanism
relating the appearance of dendrites in dielectric breakdown\cite{vera}
and in viscous fingers, we began to explore the idea of trying to
obtain the return branch that is typically found in natural lightning,
in simple viscous fingers experiments. In a normal viscous finger
experiment using a radial Hele-Shaw cell, a low viscosity fluid (water
for example) is injected from the center towards a high viscosity
fluid that fills the gap between the upper and lower plates. This
setup is not favorable for our purposes, because if we put water at
the outer boundary, the continuous outwards flux of the high viscosity
fluid will prevent water to advance against this flux. Some years
ago people began to explore the effects of using more complex fluids
in viscous fingers experiments\cite{McCloud-Maher1995}. Experiments
where the high viscosity fluid was replaced by foams revealed dendritic
growth instead of soft viscous fingers, with structures persisting
in the limit of zero velocity. Because the dendrites obtained in experiments
using foams are similar to the structures obtained in dielectric breakdown,
and because it is possible to eliminate the outwards flux, this kind
of experiment is adequate for our purposes. It was not clear that
we could observe the appearance of a return branch in experiments
using foams, because in real lightning the main branch measures several
hundred meters and the return branch is in the order of thirty meters.

We constructed our radial Hele-Shaw cell using a 44 x 44 cm acrylic
tray with raised edges of 5 cm to contain two circular plates and
the low viscosity fluid, that will be present at the outer boundary
of the plates (see fig. 1). A circular glass plate 0.8 cm thick and
38 cm in diameter was glued to the bottom of the tray, a similar glass
plate with a 0.8 mm hole at its center was put on top of the lower
plate and was fixed to the bottom of the tray by four handles, a small
cylindrical container for the low viscosity fluid was attached to
the central hole and connected to a constant pressure source. The
main characteristics of this setup are: to allow the presence of the
low viscosity fluid at the boundary, a clean boundary because the
handles are attached to the upper side of the top glass plate, the
surfaces in contact with the foam are flat and rigid. We used four
calibrated spacers to set the separation between the plates, then
we adjusted the four handles, removed the spacers and fixed the handles.

\begin{figure}
\includegraphics[%
  width=1.0\columnwidth]{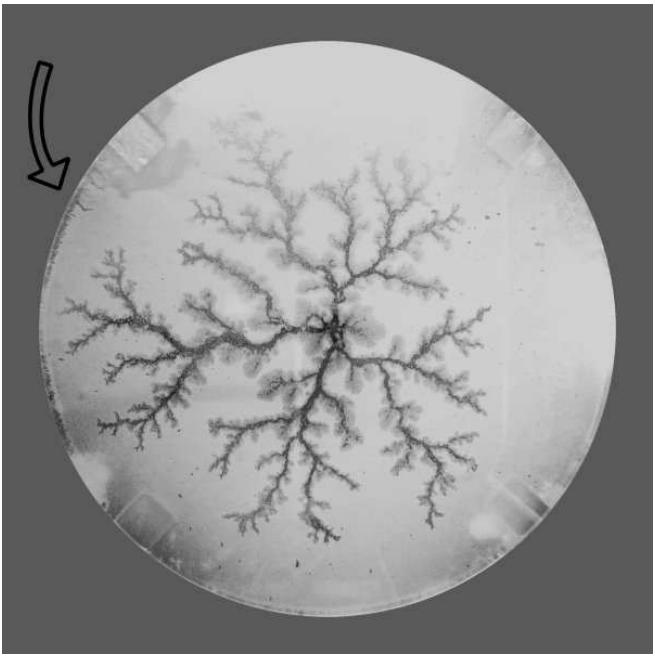}

\caption{Dendrites obtained in our experiments showing the initial stages
of formation of a return branch.}
\end{figure}

Figure 2 shows a pattern obtained injecting colored water from the
central hole, the circular boundary is under water at atmospheric
pressure, the gap between the plates was set to a separation of 0.2
mm and filled with a commercial shaving foam, a constant pressure
of 47 mm of Hg was used in this experiment. This pattern is similar
to that shown in figure~1 of ref~\cite{pietronero1984} corresponding
to a leader discharge in a gas confined between two plates, the fractal
dimension\cite{mandelbrot1977} measured for that experiment was $D=1.7$.
During this experiment we obtained a series of high resolution images
using a digital camera, in the image shown, we have cut the region
outside the circular boundaries of the plates and filled that region
with a gray color, the inner region was converted to gray levels and
only the brightness and contrast were modified.

Using the previous parameters, a typical experiment lasts for about
thirty seconds, this slow evolution makes very easy to observe how
patterns grow and to capture images with a camera. When the left branch
shown in fig. 2 is near the boundary, water from the low pressure
region begin to invade the foam, but as this branch advances towards
the boundary it swept the foam outwards, preventing the return branch
to be formed. Nevertheless, in the upper region (see the arrow) near
the tip of this branch, there is clear evidence for the appearance
of the return branch, in this region and in the lower region near
the tip, one can see the water that invaded the foam. The effect is
more evident when one analyze the sequence of high resolution images,
advancing forwards and backwards in time makes clear the two competing
effects: water from the low pressure region invades the foam and the
foam is swept outwards by the advancing main branch. To obtain
the return branch, it is crucial to eliminate this
second effect. By this reason, we used water and foam and not water
and oil as in a normal viscous finger experiment. We expect that by
using other kind of foams we could make the effect more impressive
and more similar to the equivalent upward-connecting leader of lightning,
we have tried many separations between the plates and several combinations
of water, oil, gels and white of egg, with no success.

The return branch is very easy to produce, in fact at early stages
of our experiment we used two acrylic plates instead of the glass
plates, and the return branch was formed under similar conditions.
This return branch induce to think that there is an universal mechanism
that governs lightning and viscous fingers. The appearance of the
return branch in this experiment is important because it shows universality,
but it also goes against intuition, water from the outer boundary
advances against the pressure field to meet the main branch. Our explanation
for the appearance of this return branch, is that the system is lowering
its energy by releasing the stress, and the return branch is a local
effect that allow the system to lower its energy globally. Although
this experiment was motivated by numerical results from a model of
one of the authors\cite{vera}, we have not included that model in
this paper to explain the appearance of the return branch, because
the theory behind the model is controversial.

We used thick glass plates to eliminate the possibility of bending.
To be sure that the return branch was not caused by an induced separation
of the plates, at late stages of some experiments we applied an upwards
force to a point near the boundary of the upper plate, this force
was similar in strength to the force driving the constant pressure
source used in our experiments, no effect on the already formed pattern
or in the water at the boundary was observed.

\end{document}